\documentclass[
    ,final            
  ]
  {aipproc}

\layoutstyle{6x9}

\begin{document}

\title{Polarization and asymmetries in neutral strange particle production}


\author{Andrew Cottrell \newline for the ZEUS Collaboration}{
  address={Denys Wilkinson Building, Department of Physics, Oxford, OX1 3RH}
}

\begin{abstract}
Inclusive $\Lambda$, $\bar{\Lambda}$ and $K^0_s$ production in deep inelastic $ep$ 
scattering has been studied with the ZEUS detector at HERA using an integrated 
luminosity of $120 \rm{pb}^{-1}$. Differential cross sections, baryon to antibaryon 
production asymmetry and baryon to meson production ratios have been measured in 
the laboratory system for $Q^2 > 25 \rm{GeV}^2$. 
\end{abstract}

\maketitle


\section{Introduction}

This paper discusses the production of the neutral strange particles $\Lambda$, $\bar{\Lambda}$\footnote{Hereafter,both $\Lambda$ and $\bar{\Lambda}$ are referred to as $\Lambda$, unless explicit comparison are made.} and $K^0_s$ in deep inelastic scattering at HERA.  As well as measuring differential cross sections for their production, the baryon-antibaryon asymmetry and the baryon to meson ratio are investigated, and a first ZEUS measurement of the transverse and longitudinal $\Lambda$ polarization is made.  

All of these measurements endeavour to clarify in different ways the transition from quark to hadron.  For example the $\Lambda$ to $K^0_s$ ratio has been measured in $e^+ e^-$ colliders \cite{PDBook} and in heavy ion collisions \cite{rhic_ratio}.  The ZEUS measurement in $ep$ collisions adds more information in trying to understand when a baryon is produced, and when a meson.  

$\Lambda$ transverse polarization also gives information on hadron production.  The DeGrand-Miettinen model \cite{degrand} explains the $\Lambda$ spin as being carried predominantly by the s quark which picks up polarization via Thomas Precession when it is accelerated.  Hence a measurement of the transverse polarization will give information on the initial direction of the s quark that ends up in the $\Lambda$.  

Observing these particles also allows an investigation into baryon number and how it is transported.  Significant baryon number
  transport over several units of rapdidity has been observed in heavy ion
  collisions \cite{rhic_baryon}.  Various models have been developed to explain this, including associating baryon number with valence quarks and moving it through rapidity by multiple scattering \cite{rhic_valence} or associating baryon number with a gluonic junction \cite{rhic_gluon}.  In HERA $ep$ collisions initially a baryon number of +1 exists as the proton moving down the beampipe.  The possibility of this baryon number being observed in the $\Lambda$ system in the central rapidity region is investigated.

\section{Event Selection and Analysis}
This analysis uses an inclusive sample of neutral current deep inelastic scattering (DIS) events collected by ZEUS in the 1996-2000 HERA running period, corresponding to an integrated luminosity of 120 $\rm{pb}^{-1}$.  The kinematic region was $Q^2> 25 \rm{GeV}^2$ and $0.02 < y < 0.95$.  

$\Lambda$ and $K^0_s$ are detected in the $p\pi$ and $\pi^+ \pi^-$ decay channels 
respectively.  A secondary vertex is observed and the mass is reconstructed from the momenta of
two oppositely charged tracks coming from the vertex.  Both tracks are assumed to have the mass of a $\pi^+$ (for 
$K^0_s$) or the track with the most momentum has the proton mass and the 
other the mass of a $\pi^+$ (for $\Lambda$).  Combinatorial background is 
removed with a bin-by-bin sideband subtraction method.  

The $\Lambda$ polarization is measured via the angular distribution of the decay products:
\begin{eqnarray}
\frac{dN}{d\Omega} \propto \frac{1}{4\pi}(1 \pm \alpha P \cos\theta)
\end{eqnarray}
in the $\Lambda$ rest frame, where $\alpha = 0.642 \pm 0.013$\cite{PDBook} is the decay asymmetry parameter and 
$P$ is the polarization.  $\theta$ is the angle between the decay proton momentum, $\vec{p}$ 
and the $\Lambda$ momentum, $\vec{P}_{\Lambda}$ (longitudinal polarization) or between 
$\vec{p}$ and $\textbf{n}$ = $\vec{P}_{beam} \times \vec{P}_{\Lambda}$, where $\vec{P}_{beam}$ is the momentum of the electron beam (transverse polarization).

\section{Monte Carlo Simulation}
Data were corrected to hadron level by using the ARIADNE 4.08\cite{cpc:71:15} Monte Carlo (MC) interfaced to HERACLES
via DJANGOH 1.1\cite{spi:www:djangoh11}. The parton density functions were
taken from the CTEQ4D set. 
The strange suppression factor, $\lambda_s$ was set to 0.3.  
ARIADNE is based on the Color Dipole Model and the LUND string model\cite{prep:97:31} is used to simulate
the fragmentation of the partons.
The Ariadne prediction of the cross sections is also shown on the results plots.  

\section{Results}

\begin{figure}
  \includegraphics[height=7cm]{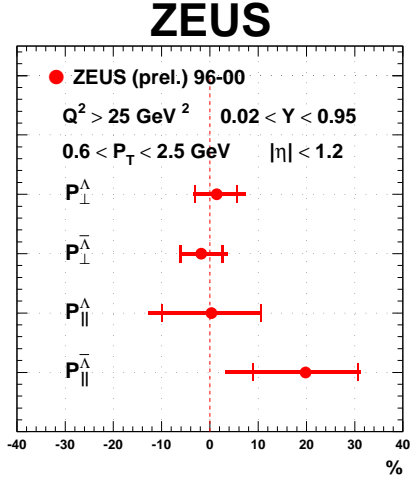}
  \caption{Transverse and longitudinal $\Lambda$ and $\bar{\Lambda}$ polarization. \label{poln} }
\end{figure}

The results of the $\Lambda/\bar{\Lambda}$ polarization are presented in Fig. \ref{poln}. Averaged over
the full kinematic range, the transverse polarization of $\Lambda$ and $\bar{\Lambda}$ were observed to be
$+1.4\pm4.5$(stat)$^{+4.1}_{-1.9}$(syst)\% and $-1.8\pm4.4$(stat)$^{+3.1}_{-1.3}$(syst)\% respectively. 
The longitudinal polarization of $\Lambda$ and $\bar{\Lambda}$ was also measured to be consistent with zero within the total uncertainties.  The longitudinal polarization was observed to be $+0.3\pm10.3$(stat)$^{+0.1}_{-7.8}$(syst)\% for $\Lambda$ and
$+19.8\pm10.8$(stat)$^{+4.2}_{-12.6}$(syst)\% for $\bar{\Lambda}$.  

\begin{figure}
  \includegraphics[height=9cm]{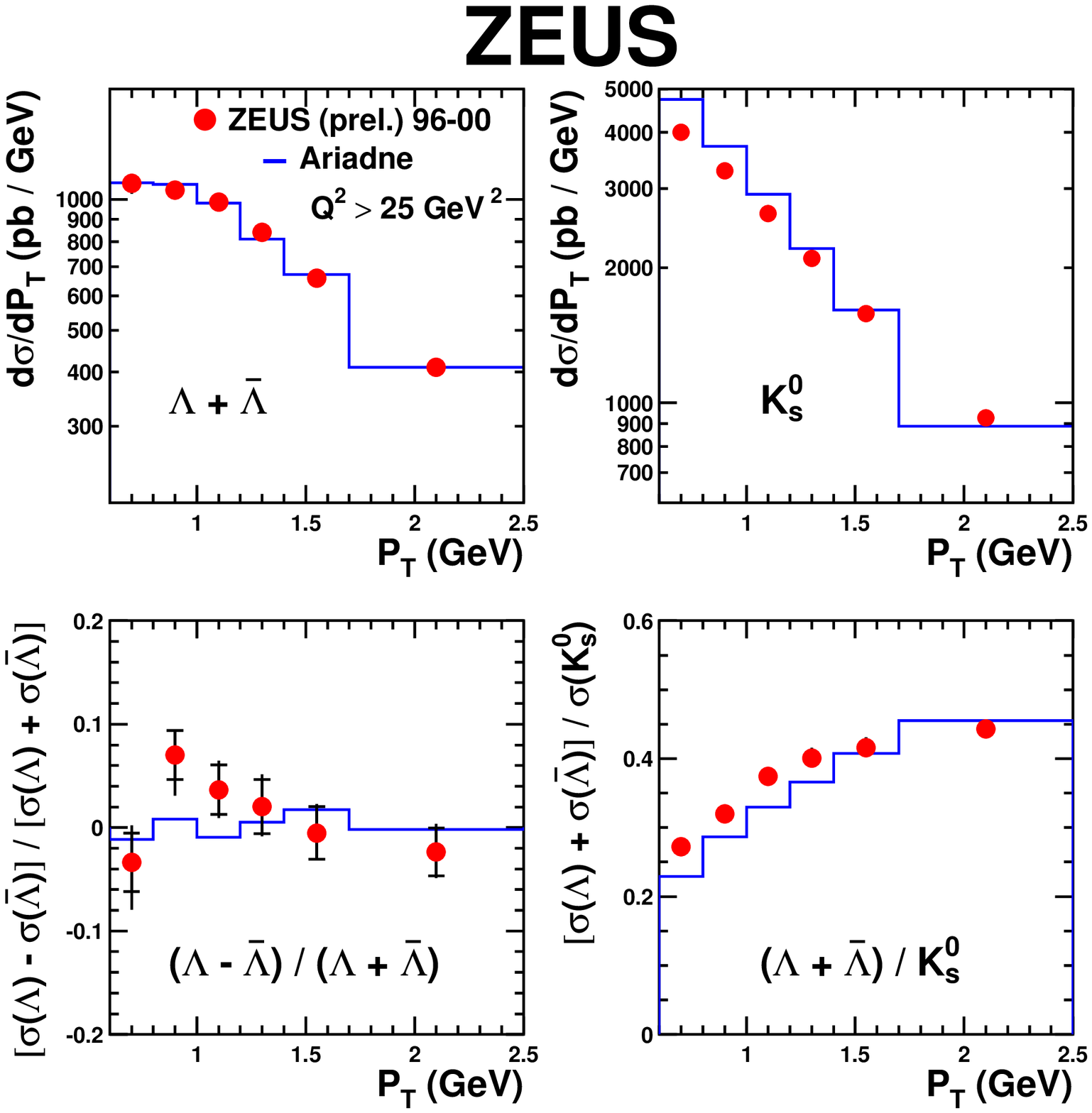}
  \includegraphics[height=9cm]{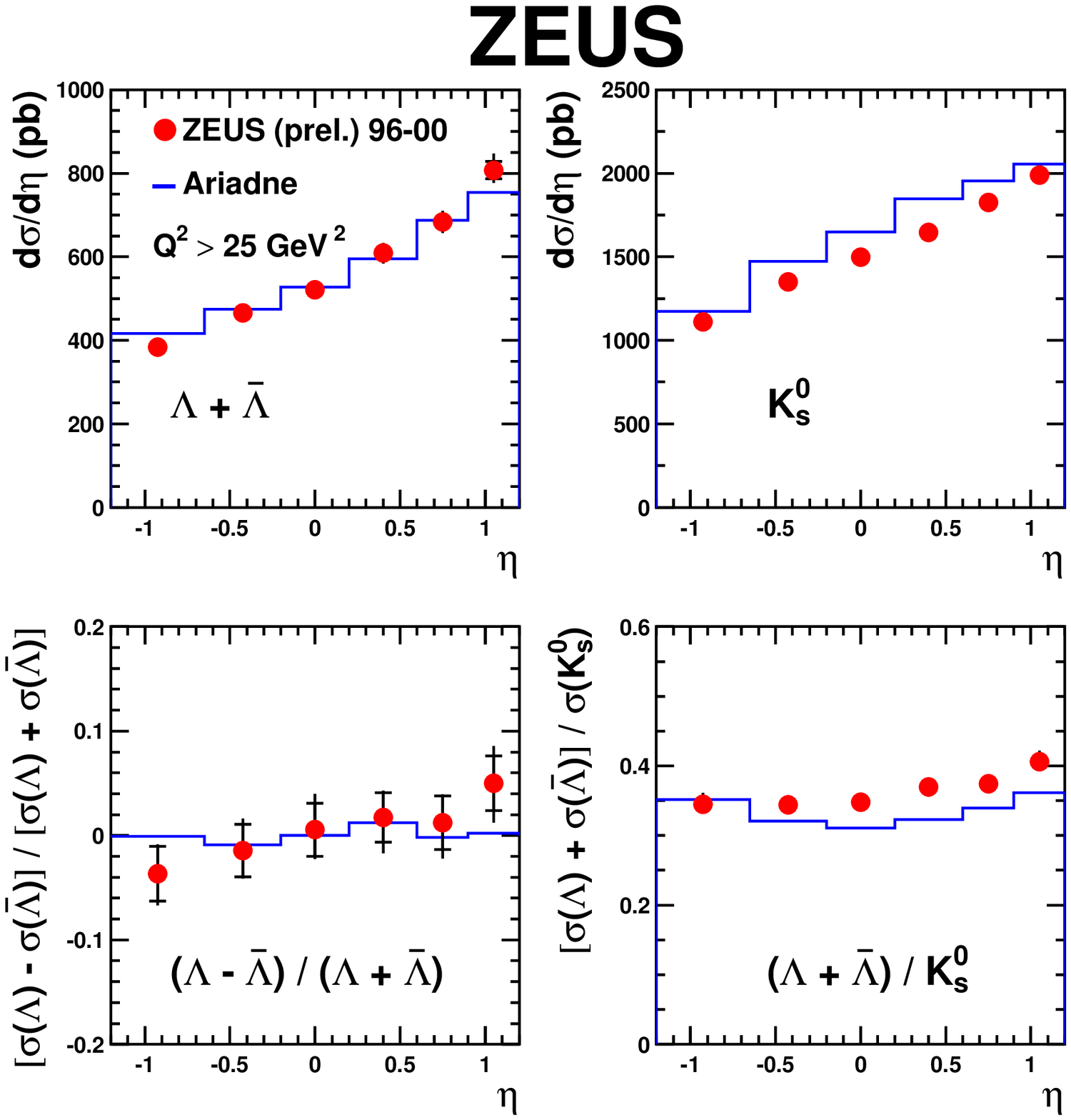}
  \caption{$\Lambda$ and $K^0_s$ cross sections, $\frac{\Lambda - \bar{\Lambda}}{\Lambda + \bar{\Lambda}}$ ratio and $\frac{\Lambda + \bar{\Lambda}}{K^0_s}$ ratio as a function of $P_T$ and $\eta$ \label{strange_pteta} }
\end{figure}

\begin{figure}
  \includegraphics[height=9cm]{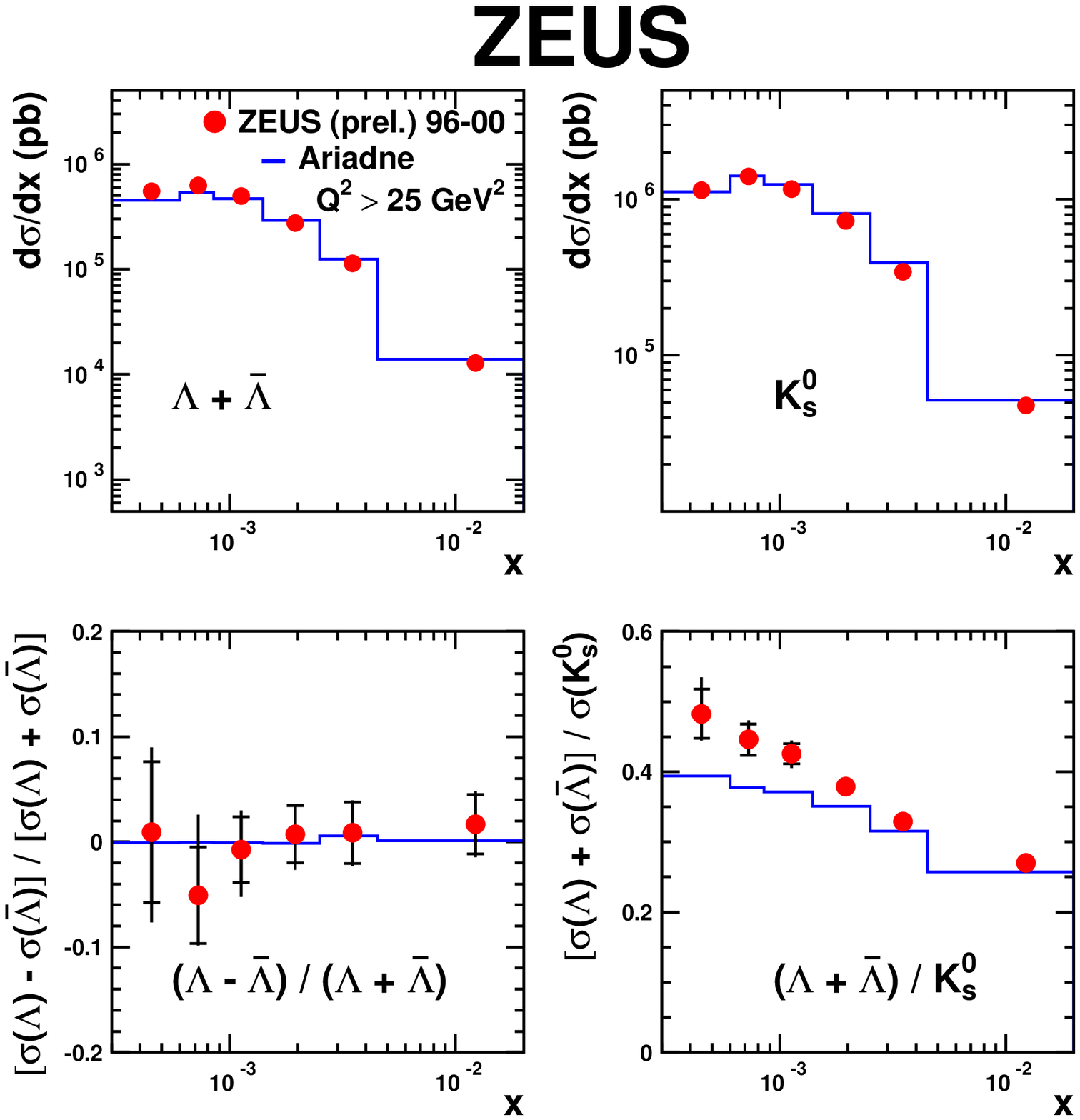}
  \includegraphics[height=9cm]{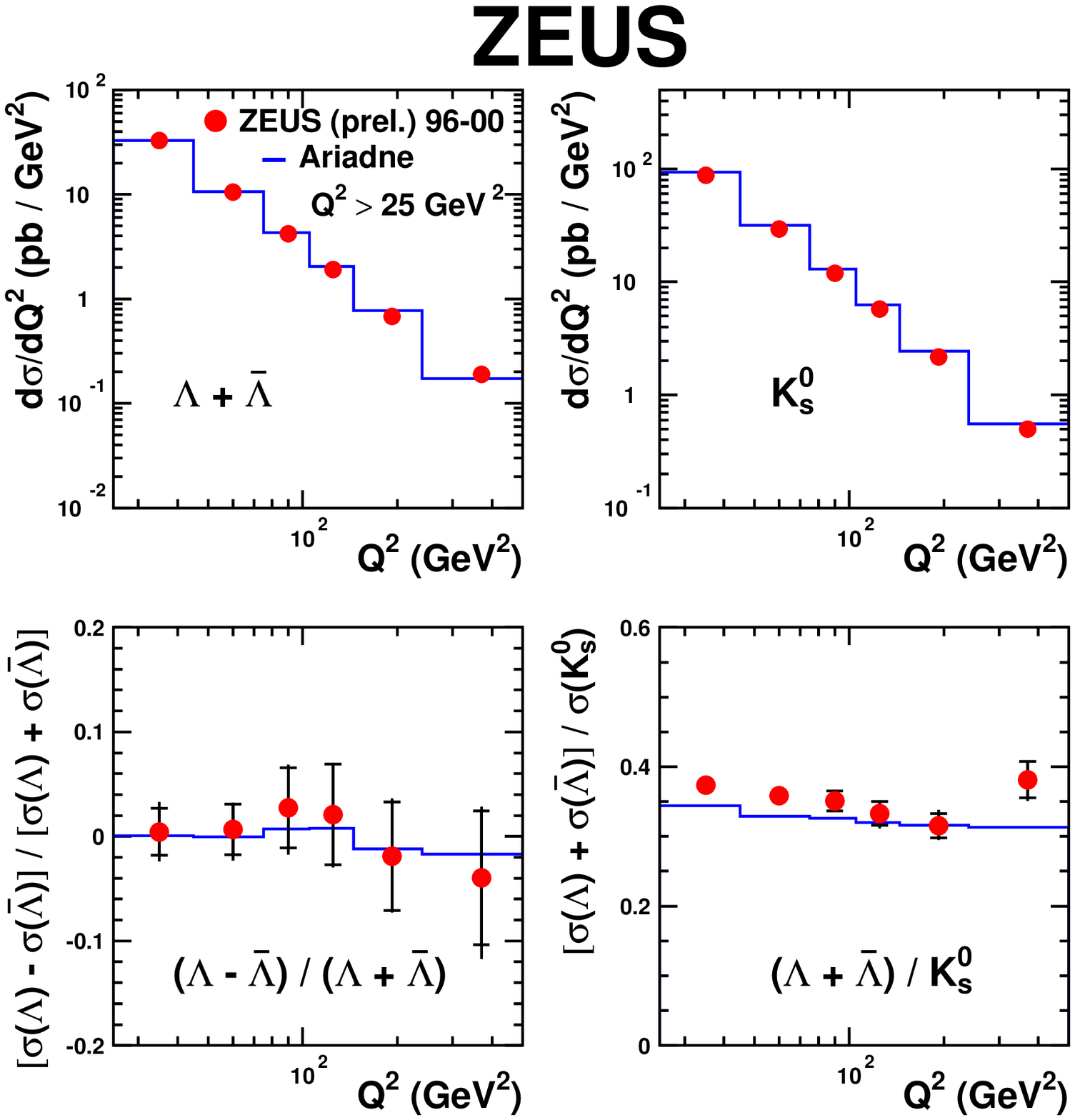}
  \caption{$\Lambda$ and $K^0_s$ cross sections, $\frac{\Lambda - \bar{\Lambda}}{\Lambda + \bar{\Lambda}}$ ratio and $\frac{\Lambda + \bar{\Lambda}}{K^0_s}$ ratio as a function of $x$ and $Q^2$ \label{strange_xq2}}
\end{figure}

The differential $\Lambda$ and $K^0_S$ cross sections, $\frac{\Lambda - \bar{\Lambda}}{\Lambda + \bar{\Lambda}}$ 
asymmetry and $\Lambda/K^0_S$ ratio as a function of $p_T$, $\eta$, $x$ and $Q^2$ are shown in Fig. \ref{strange_pteta} and Fig. \ref{strange_xq2}.
The MC generally describes the cross sections in the data.  
However, the MC tends to overestimate the $K^0_s$ production, particularly at low 
$P_T$.  

The ratio $\frac{\Lambda - \bar{\Lambda}}{\Lambda + \bar{\Lambda}}$ as a function of all kinematic variables is consistent with zero.  

The $\Lambda/K^0_S$ ratio is generally described by the MC.  However, although the steep rise as $x$ decreases is modelled by the MC to some extent, it is not to the same degree.  Other differences between the MC and the data can be seen as a function of $P_T$, where the $\Lambda/K^0_s$ ratio is in excess of the MC at low $P_T$, and as a function of $\eta$ where the MC is symmetric about $\eta = 0$, whereas the data shows an increase in this ratio as $\eta$ increases.

\section{Conclusions}
Using the model of Degrand and Miettinen \cite{degrand}, the lack of any 
transverse polarization suggests that the strange quarks in the the $\Lambda$ baryons 
observed by ZEUS do not come from any particular direction.  The longitudinal 
polarization being consistent with zero is as expected with HERA-I data, but 
gives a measure of the potential to measure any polarization transfer from the 
electron beam to the $\Lambda$ in HERA-II.  
The $\Lambda/K^0_S$ ratio has been measured, and the data is generally described 
by ARIADNE.  Areas of the phase space do exist where the 
MC is not sufficient to describe the data, particularly at low $P_T$ and 
low $x$.  No significant $\Lambda - \bar{\Lambda}$ asymmetry is seen.





\bibliographystyle{aipproc}   

\bibliography{dis_lam}

\hyphenation{Post-Script Sprin-ger}
\begin{thebibliography}{9}
\expandafter\ifx\csname natexlab\endcsname\relax\def\natexlab#1{#1}\fi
\providecommand{\enquote}[1]{``#1''}
\expandafter\ifx\csname url\endcsname\relax
  \def\url#1{\texttt{#1}}\fi
\expandafter\ifx\csname urlprefix\endcsname\relax\def\urlprefix{URL }\fi
\providecommand{\eprint}[2][]{\url{#2}}

\bibitem[{K. Hagiwara }{ et al.}(2002)]{PDBook}
{K. Hagiwara }{ et al.}, \emph{{Phys. Rev. D}}, \textbf{66}, 010001+ (2002),
  \urlprefix\url{http://pdg.lbl.gov}.

\bibitem[{Huang} and {Rafelski}(2005)]{rhic_ratio}
H.~{Huang}, and J.~{Rafelski}, \emph{{hep-ph/0501187}}, 2005.

\bibitem[{DeGrand} and {Miettinen}(1981)]{degrand}
T.~{DeGrand}, and H.~{Miettinen}, \emph{{Phys. Rev. D}}, \textbf{24}, 2419
  (1981).

\bibitem[{S. Bass}{ et al.}(2003)]{rhic_baryon}
{S. Bass}{ et al.}, \emph{{Phys. Rev. Lett. }}, \textbf{91}, 052302 (2003).

\bibitem[{S. Bass}{ et al.}(2004)]{rhic_valence}
{S. Bass}{ et al.}, \emph{{J. Phys. G: Nucl. Part. Phys.}}, \textbf{30}, 1283
  (2004).

\bibitem[{S. Vance}{ et al.}(1998)]{rhic_gluon}
{S. Vance}{ et al.}, \emph{{Phys. Lett. B}}, \textbf{443}, 45 (1998).

\bibitem[L\"onnblad(1992)]{cpc:71:15}
L.~L\"onnblad, \emph{{Comp. Phys. Comm.}}, \textbf{71}, 15 (1992).

\bibitem[Spiesberger(1998)]{spi:www:djangoh11}
H.~Spiesberger, \emph{{\em {\sc heracles} and {\sc djangoh}: Event Generation
  for $ep$ Interactions at {HERA} Including Radiative Processes}}, 1998,
  \urlprefix\url{http://www.desy.de/~hspiesb/djangoh.html}.

\bibitem[{B. Andersson}{ et al.}(1983)]{prep:97:31}
{B. Andersson}{ et al.}, \emph{{Phys. Rep.}}, \textbf{97}, 31 (1983).

\end{thebibliography}

\IfFileExists{\jobname.bbl}{}
 {\typeout{}
  \typeout{******************************************}
  \typeout{** Please run "bibtex \jobname" to optain}
  \typeout{** the bibliography and then re-run LaTeX}
  \typeout{** twice to fix the references!}
  \typeout{******************************************}
  \typeout{}
 }

\end{document}